\documentstyle[a4,12pt,amssymb,epsfig,amsfonts]{article}

\def\e{{\rm e}}

\def\d{\partial}
\def\l{\left(}
\def\r{\right)}

\newcommand{\be}{\begin{equation}}
\newcommand{\ee}{\end{equation}}

\renewcommand{\ln}{\mathop{\rm ln}\nolimits}

\newcommand{\bg}{\begin{gather}}
\newcommand{\eg}{\end{gather}}

\begin{document}
\begin{center}
{\Large \bf
On electric charge non-conservation \\
in brane world}
\\ 
\vspace{0.3cm}
S.L.~Dubovsky and V.A.~Rubakov\\
{\small{\em 
Institute for Nuclear Research of the Russian Academy of Sciences, }}\\
{\small{\em
60th October Anniversary prospect 7a, Moscow 117312, Russia\\
}}
\end{center}
\begin{abstract} 
In some models with  infinite extra dimensions, gauge fields
are localized on a brane by gravity.
A generic property of these
  models is the existence of arbitrarily light bulk modes of charged
  fields. This property may lead to interesting
  low-energy effects such as electric charge non-conservation
on the brane (decay of electron to nothing). 
  One may worry that light charged Kaluza--Klein
  modes would lead to 
  unacceptable phenomenology due to their copious production in, e.g.,
  electron-positron annihilation
and/or their contribution to QED observables like anomalous
magnetic moments. We argue, however, that both loop effects and
  production of light charged
  Kaluza--Klein modes are suppressed due to  screening effect of
  gapless spectrum of bulk photons.
\end{abstract} 
An idea that we may live on a domain wall in a multi-dimensional
space~\cite{RS,Akama:1982jy} has been revived recently, with
emphasis on gauge hierarchy 
problem~\cite{Arkani-Hamed:1998rs,Randall:1999ee}. 
A particularly noteworthy development of the brane world idea is
the possibility to localize gravitational field on a brane
in a theory with infinite extra dimension~\cite{LRS}; this became an
alternative to the Kaluza--Klein (KK) compactification
(see Ref.~\cite{Rubakov} for a review).

A phenomenologically viable implementation
of the latter idea should involve a mechanism of the
localization of gauge fields on a brane. While a number of field
theoretical models aimed to solve this problem were 
proposed~\cite{Dvali:1997xe,Dvali:2001rx,Akhmedov:2001ny} 
(see also Ref.~\cite{Dubovsky:2001pe}
for  discussion of general properties of such models), it is tempting
to use gravitational field of the brane itself to localize gauge
fields, thus providing a universal mechanism for localization of gauge
fields and gravity.

While there is no localized gauge field in the original
Randall--Sundrum model, one can slightly modify this setup by adding
extra warped compact dimensions to obtain a normalizable photon zero
mode~\cite{Oda:2000zc,DRT2}. Due to this mode, 
interactions of charged particles residing on the
brane are described, at the
classical level, by  four-dimensional Maxwell's electrodynamics, up
to small power-law corrections suppressed by the ratio of energy
to the  
inverse adS radius $k$.

One may expect a number of new effects in this setup.
Indeed, a generic property of models with warped infinite extra
dimensions is that the spectrum of bulk modes of all fields
starts from 
zero four-dimensional mass irrespectively of the mass of the localized
mode. As a result, massive brane particles in these models are
metastable resonances in the continuous spectrum of bulk
modes~\cite{DRT1}. Decay of such a resonance, {\it i.e.}
escape of a  particle from the brane, would show up,
  from the point of view of an observer on
the brane,
as literal
disappearance of the particle.
Of  particular interest from both theoretical and
phenomenological points of view would be a possibility for 
{\it charged}
particles to escape from the brane. 
The processes
like decay $e^-\to nothing$ would be  a convincing signature
of 
 infinite extra dimensions~\cite{DRT2}.

This proposal may appear controversial from both  classical and
quantum points of view. On the classical side, causality and four-dimensional
Gauss' law appear to forbid electric charge non-conservation.
This apparent discrepancy
 is resolved by the fact that  electric charge is
conserved in the complete multi-dimensional theory. Namely,
a charged particle escaping into the bulk leaves behind
a spherical wave of 
electric field on the brane. This wave wipes out the 
four-dimensional Coulomb
field on the brane
in a causal manner~\cite{DRT2} (see also
Refs.~\cite{Gregory:2000rh,Giddings:2000ay} 
where gravitational analogue of this
phenomenon is discussed). An equivalent description of the same process,
inspired by the adS/CFT correspondence, would be to
say that the charged particle decays into  spherical cloud of charged
(quasi-)conformal matter (cf. Ref.~\cite{Giddings:2000ay}).

Problems with this scenario  appear  much
more severe at the quantum level.
These are related to the observation that
successful localization of gauge field necessarily implies charge
universality~\cite{Dubovsky:2001pe}. In particular, the
bulk modes of the charged fields 
interact with the localized photon at the same strength 
as the brane modes.
Then one may worry that the existence of light bulk modes may be
phenomenologically unacceptable because of their loop contributions
into precision QED observables and/or  their copious production 
 in various processes like, e.g,
$e^+e^-$-annihilation. 

The purpose of this note is to discuss this issue in some detail.
We will argue that it is possible to have light bulk modes of 
charged fields (so that electron disappearance  is allowed)
without disastrous loop effects or
overproduction of these modes in the processes like $e^+e^-$-annihilation.  The corresponding suppression is due to
the continuum of bulk photon modes which ``screens'' the
interaction between charged bulk modes and charges residing on the
brane.

To be specific, let us consider a model for the localization of gauge
fields suggested in Refs.~\cite{Oda:2000zc,DRT2}. One begins with
the solution to ($4+n+1$)-dimensional Einstein equations,
\begin{equation}
\label{*}
ds^2={1\over (1+k|\xi|)^2}
\left(dt^2-d{\bf x}^2-\sum_{i=1}^{n}R_i^2d\theta_i^2-d\xi^2\right)
\end{equation}
where $\theta_i\in [0,2\pi]$ are compact coordinates,  $R_i$ are 
radii of compact dimensions and
$k$ is the inverse adS radius determined by the bulk 
cosmological constant. 
There is a single  brane  located at $\xi =0$.
The only difference between this metric and the
Randall--Sundrum metric is the
presence of  extra compact dimensions $\theta_i$. These dimensions
are added for obtaining a localized zero mode of the gauge field.
In what follows we assume that their radii $R_i$ are
the smallest length scales involved, so all fields are
taken independent of $\theta_i$. The inverse adS radius $k$ is assumed
to be the largest energy scale involved.

Let us now introduce the $U(1)$ gauge field propagating in the 
background metric (\ref{*}). With appropriate rescaling, the effective
action for $\theta$-independent gauge field is
\be
\label{photon}
S=-{1\over 4g^2}\int {d\xi d^4x\over (1+k|\xi|)^{n+1}}
F_{\alpha\beta}F^{\alpha\beta}\;, 
\ee 
where indices are raised and lowered
by the metric of  flat $(4+1)$-dimensional
Minkowski space-time. We
assume that there are no currents along compact dimensions and disregard
$A_{\theta_i}$ components of the vector field, so that
$\alpha,\beta=0,\dots,3,\xi$. 

 As discussed
in Refs.~\cite{Oda:2000zc,DRT2}, for $n > 0$ there exists
 a localized zero mode
\[
A^{(0)}_{\mu}(\xi)=const
\]
($\mu=0,\dots,3$).  Also, there is a
continuous spectrum of  bulk modes with four-dimensional
masses $m$ starting from
zero. 

To study (apparent) charge non-conservation in this setup we
 introduce light charged fields both on the brane  and
in the bulk. One  way to do this~\cite{RS,Bajc}
is to introduce a
single charged fermion field $\psi$. One finds the following 
effective action
for the $\theta$-independent field,
\be
\label{yukawa}
S_{ferm}=\int d\xi d^4x\sqrt{|g|}\bar{\psi}\l
i\gamma^\alpha \nabla_\alpha
-{k(n+4)\over 2}\gamma^5\mbox{sign}(\xi)+y\phi_k(\xi)\r\psi\;,
\ee
where $\nabla_\alpha$ includes spinor connection corresponding to 
metric (\ref{*}) and minimal interaction 
with the photon field $A_\alpha$. 
A term ${kn\over 2}\gamma^5\mbox{sign}(\xi)$ 
comes from the
$\theta_i$ components of spinor connection. To
localize the fermion on the brane, we have added  
Yukawa-type  interaction
$y\phi_k(\xi)\bar{\psi}\psi$  with the
background scalar field $\phi_k(\xi)$, having  kink profile,
\[
\phi_k(\xi)=v\mbox{sign}(\xi)\;.
\]
It is convenient
to introduce  rescaled fermion field 
\[
\chi(\xi)\equiv(1+k|\xi|)^{n/2+2}\psi(\xi)\; .
\]
The action
for the rescaled field takes the following simple form
\be
\label{ferm}
S_{ferm}=\int d\xi d^4x\left(\bar{\chi}i\gamma^{\alpha}{\cal
D}_{\alpha}\chi+{yv\mbox{sign}(\xi)
\over
1+k|\xi|}\bar{\chi}\chi\right)
\ee
where ${\cal D}_{\alpha}$ is  flat space covariant derivative,
which includes the field $A_{\alpha}$. 

It is straightforward to check that a left-handed fermion 
zero mode localized
on the brane and  gapless continuum spectrum are present in this
setup~\cite{Bajc,DRT1}. 
Following Ref.~\cite{DRT1} one can introduce  small mass for
the localized mode;
the continuum will still start from zero whereas the massive brane
particle will decay into
light continuum modes, {\it i.e.} escape from the brane. 
From the point of view of 
four-dimensional observer,
the latter process would show up as
(apparent) charge non-conservation.

One remark is in order.
The prefactor $g^2(1+k|\xi|)^{n+1}$ in Eq.~(\ref{photon})
may be viewed as
the $\xi$-dependent coupling constant. It increases towards large
$|\xi|$, so once the charged bulk fields are introduced,
the theory becomes strongly coupled at large $|\xi|$.
 We proceed under assumption that the overall features of the spectrum
remain unaffected by gauge interactions: there are light bulk fermions
which are capable of moving towards large $|\xi|$ and whose mixing
with brane electrons makes the latter unstable against the escape into
extra dimension.

The problem we would like to discuss exists already in
the 
massless case. Namely, if one performs 
Kaluza--Klein decomposition of the fermion field, one finds that
interactions of the photon zero mode with the localized fermion mode and
with fermion
modes from the continuum  are of  equal strength. A
clear way to see this is to put a system in a large box of 
size $L$ in the $\xi$-direction. Then the effective charge $g_{eff}$
describing the interaction between the photon zero mode and a
Kaluza--Klein mode with the wave function $\chi(\xi)$ is proportional
to the overlap integral
\[
g_{eff}\propto
\int_{-L}^Ld\xi A_{\mu}^{(0)}(\xi)\bar{\chi}(\xi)\chi(\xi)\;.
\]
In our case, the
 photon zero mode $A_{\mu}^{(0)}(\xi)$ is constant, so this
integral reduces just to the normalization integral for the wave
function $\chi(\xi)$ and leads to one and the same value of the effective
charge irrespectively of the profile of the wave function
$\chi(\xi)$ (charge universality). 
Were this the whole story, one would face  complete
phenomenological disaster: besides the zero fermion mode aimed to describe
the four-dimensional electron there is a plethora of arbitrarily light
particles (Kaluza--Klein modes) of the same charge. 
The latter would make strong contributions to precision
QED observables and would be 
copiously produced in
the processes like $e^+e^-$-annihilation.

However, there is also  gapless continuum spectrum of photon bulk
modes. The corresponding wave functions are small at $\xi=0$, but grow
towards large $|\xi|$. Consequently, these modes interact
weakly with the localized fermion, but their interaction with bulk
fermion modes is quite strong. The main observation of this note is that
in a certain regime, the latter interaction screens the effect of the
photon zero mode, thus leading to  phenomenologically acceptable
results.

To be specific, let us study the annihilation of $e^+e^-$-pair on the brane
into bulk modes (the issue of loop effects is simpler, as we discuss
in due course). From the above discussion it is clear that the 
Kaluza--Klein
decomposition is not an adequate tool to study this process, as the
couplings between bulk photons and bulk fermions  are badly IR divergent
due to the growth of the corresponding wave functions as
$|\xi|\to\infty$. A way to soften this ``strong coupling'' problem is to
work with propagators in the $\xi$-coordinate representation 
instead of performing the Kaluza--Klein
decomposition (cf. Ref.~\cite{strongcoupling,LRMS}).

To avoid unnecessary technical difficulties and to make the underlying
physics more transparent we   simplify our setup further. First,
instead of the bulk vector field $A_\alpha$ we consider  scalar field
$A$. Further, instead of considering the bulk charged fermion field we
introduce  an additional (tensionless) brane at $\xi=\xi_0$ and assume
that the field content of our toy model consists of the scalar bulk
field $A(x,\xi)$ (``photon''), fermion field $e(x)$ (``electron'')
living on ``our'' brane at $\xi=0$ and fermion field  $b(x)$
(``brother  electron'') living on
the second brane at $\xi=\xi_0$. Then the action takes the
following form
\begin{eqnarray}
S=\int d^4xd\xi\left[{1\over
2g^2(1+k|\xi|)^{n+1}}(\d_{\alpha}A)^2+\delta(\xi)\bar{e}i\gamma^\mu\d_{\mu}e+
\delta(\xi-\xi_0)\bar{b}i\gamma^\mu\d_{\mu}b\right.\nonumber\\
\label{6*}
\left.+\delta(\xi)A\bar{e}e+
\delta(\xi - \xi_0)A\bar{b}b\right]
\end{eqnarray}

It is instructive to study first the same model in  flat
space, $k=0$. Then the  photon zero mode is absent, so  physics
on the brane at $\xi=0$ is not four-dimensional even in the absence
of the field $b(x)$. However, one may still ask 
whether introducing $b$-particles changes physics at the $\xi=0$ brane, 
{\it i.e.} whether $b$-particles affect cross-sections of  
processes at the $\xi=0$ brane, like $e^+e^-$-annihilation into the
bulk 
or
into $e^+e^-$. 

Physically, due to locality, 
it is clear that one should not see the
new particles in ``hard'' processes with characteristic momenta $p\gg
\xi_0^{-1}$. To make this statement  precise, let us consider the
K\"allen-Lehmann representation for the photon propagator on the
$\xi=0$ brane,
\begin{equation}
\label{Challen}
\left.\langle T A(\xi;p)A(\xi';-p)\rangle\right|_{\xi=\xi'=0}=\int\!\!
ds\; {\rho(s)\over p^2-s}
\end{equation}
where, as usual, $\rho(s)$ is proportional to the total cross section
of $e^+ e^-$-annihilation. The
contribution of the field $b(x)$ to this propagator  comes from
the diagrams of the type shown in Fig.~1.
\begin{figure}[!h]
\label{spectral}
\begin{center}
\epsfig{file=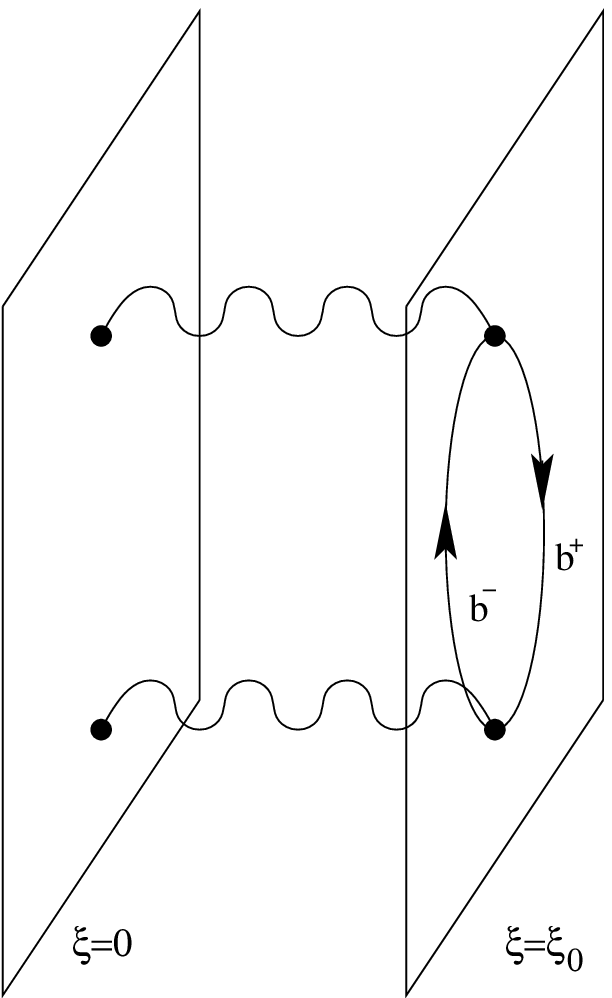,height=6.cm,width=4.cm}\\
Figure 1.
\end{center}
\end{figure}
Taking into
account the explicit form of the free propagator $G_0(\xi,\xi';p)$ for
the field $A(x)$,
\be
\label{freepropagator}
G_0(\xi,\xi';p)={1\over (2\pi)^3}{\e^{ip|\xi-\xi'|}\over2p}
\ee
we see that this contribution is exponentially small,
\be
 (b-loops) \; \propto \exp{(-2|p|\xi_0)} \; ,
\label{5a*}
\ee
 at ``large'' Euclidean momenta,
\[
|p|\xi_0\gg 1,\; p^2<0\;.
\]
This means, in the first place, that contributions of
$b$-particle loops into processes on $\xi=0$ brane are
exponentially suppressed, {\it i.e.} anomalous magnetic moments,
Lamb shift, etc., are unaffected, provided that $m_e \xi_0 \gg 1$.
This property persists in warped models with localized photon
(cf. Ref.~\cite{strongcoupling}).

The situation with the production of $b$-particles is more
subtle.
In terms of the quantity of  interest --- spectral density
$\rho(s)$ --- the
exponential suppression of the diagrams of  Fig.~1
at ``large'' Euclidean momenta
 implies that the corrections to  $\rho(s)$
due to $b$-particles have the form of rapid
oscillations
\be
\label{oscil}
\delta\rho(s)\propto a\cos(2\sqrt{s}\xi_0)+b\sin(2\sqrt{s}\xi_0)
\ee
at large values of $\sqrt{s}\xi_0$. Consequently, additional 
terms in the 
$e^+e^-$-annihilation cross-section also have the oscillating form. 
In the flat space model we discuss at the moment,
the oscillatory behavior (\ref{oscil})
is the only reason of why  $b$-particles are harmless
for large enough $\xi_0$:
rapidly oscillating contribution into the annihilation cross-section 
 is
impossible to detect provided 
the incoming electrons have an energy-momentum
spread larger than $\xi_0^{-1}$.
We will see that in the original  model with warped space and
localized photon, $e^+ e^-$-annihilation into bulk modes is
additionally suppressed.

Let us proceed with flat space model and
 illustrate the general reasoning leading to Eq.~(\ref{oscil}).
Let us 
study the diagrams of $e^+ e^-$-annihilation directly,
to the lowest orders of perturbation theory.
In the absence of the field $b(x)$, the dominant channel of
 $e^+e^-$-annihilation into the bulk is the process $e^+e^-\to
 \gamma$. This process is kinematically allowed, since there is no
 momentum conservation along the $\xi$-direction. 
The cross-section for
this process is determined by the diagram shown in
 Fig.~2
\begin{figure}[!h]
\label{feetogamma}
\begin{center}
\epsfig{file=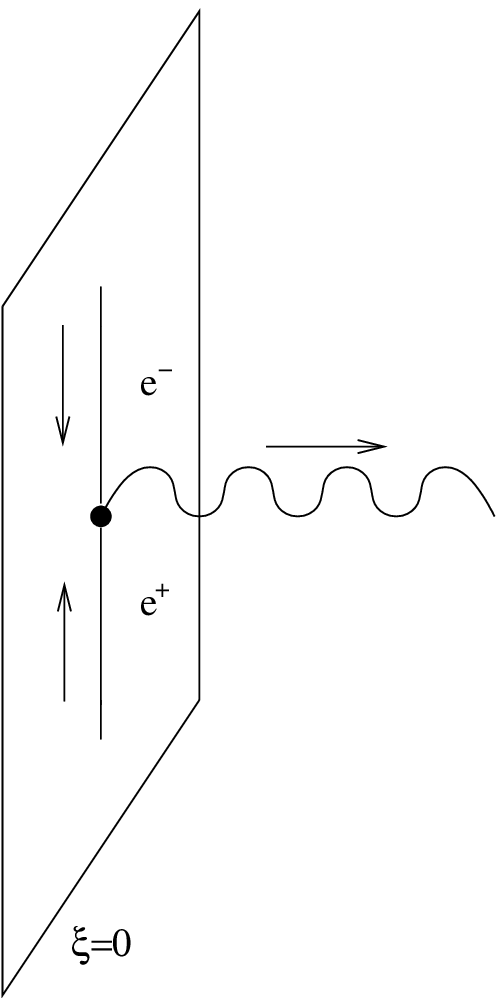,height=6.cm,width=4.cm}\\
Figure 2.\\
\end{center}
\end{figure}
and is equal to
\be
\label{eetogamma}
\sigma_0(e^+e^-\to \gamma_{bulk})=4\pi^2g^2/E
\ee
where $E$ is the center of mass energy. 
As there is no localization of the 
gauge field in  flat space, 
this cross-section is not suppressed relative
to the cross-section of $e^+e^-\to e^+e^-$-annihilation, given by
\be
\label{eetoee}
\sigma(e^+e^-\to e^+e^-)=\pi g^4/4\;.
\ee
Let us now study what happens when $b$-particles are added. Now one has
a new channel for $e^+e^-$-annihilation into the bulk --- annihilation
into the $b^+b^-$ pair. The leading order contribution to this process
comes from the diagram shown in Fig.~3
\begin{figure}[t]
\label{eetobb}
\begin{center}
\epsfig{file=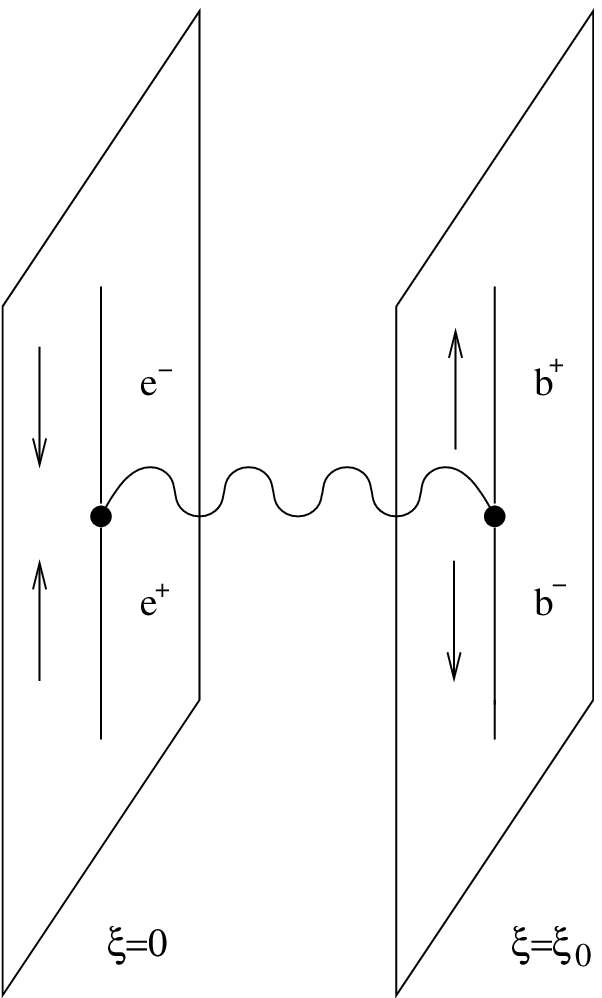,height=6.cm,width=4.cm}\\
Figure 3.\\
\end{center}
\end{figure}
 and the
corresponding cross-section is equal to the cross-section of $e^+e^-\to
e^+e^-$,
\[
\sigma(e^+e^-\to b^+ b^-)=\pi g^4/4\;.
\]
Naively, this result contradicts  our general statement  that the
correction due to $b$-particles should have 
oscillatory behavior, Eq.~(\ref{oscil}). To see what actually happens,
we recall that 
from the point of view of the observer on the
$\xi=0$ brane,   $e^+ e^-$-annihilation into $b^+ b^-$
 is indistinguishable from 
$e^+e^-\to\gamma_{{bulk}}$, as this observer cannot detect
$b$-particles or bulk photons.
Consequently, an observable
quantity is the sum of $\sigma(e^+e^-\to\gamma_{bulk})$ and
$\sigma(e^+e^-\to b^+b^-)$.
The former cross-section itself acquires loop corrections due to 
$b$-particles. Working consistently to order $g^4$, we should
take into account 
the one-loop diagram shown in
Fig.~4.
\begin{figure}[t]
\label{eetob1}
\begin{center}
\epsfig{file=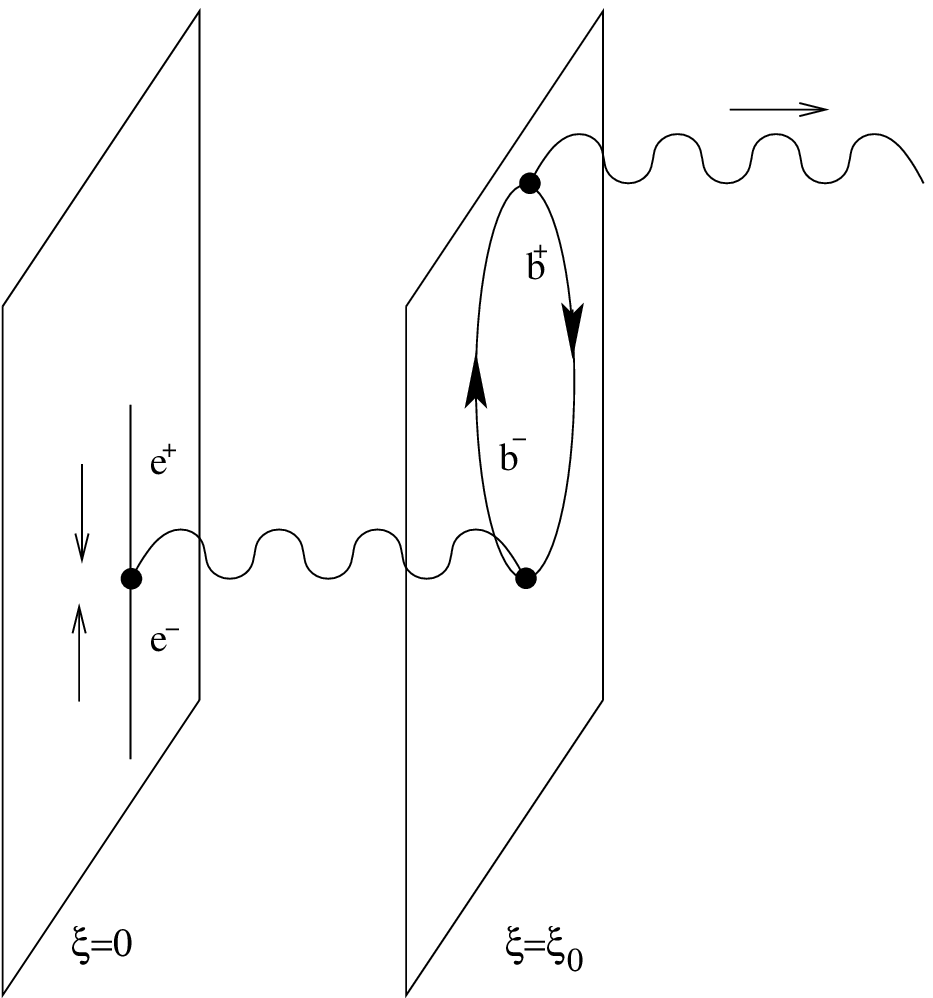,height=6.cm,width=6cm}\\
Figure 4.\\
\end{center}
\end{figure} 
It produces the energy-dependent multiplicative
``renormalization'' of the tree cross section (\ref{eetogamma}),
\begin{eqnarray}
\label{onelooptogamma}
{\sigma_{1-loop}(e^+e^-\to\gamma_{bulk})\over\sigma_0(e^+e^-\to\gamma_{bulk})}
=\;\;\;\;\;\;\;\;\;\;\;\;\;\;\;\;\;\;\;\;\;\;\;\;
\;\;\;\;\;\;\;\;\;\;\;\;\;\;\;\;\;\;\;\;\;\;\;\;\;\;\;\;\;\;\;\;\;\;\;\;\;\;\;\;\;\;\;\;\;\;\;\;
\nonumber \\
1-{g^2\over
16\pi}E-{g^2\cos(2E\xi_0)\over 16\pi}E+
{g^2\cos(2E\xi_0)\over16\pi^2}E\ln{E^2/\mu_{UV}^2}\;,
\end{eqnarray}
where we used dimensional regularization and $\mu_{UV}$ is the
normalization scale.
Here the last two (oscillating) terms are corrections to the 
cross-section to emit a photon with  negative momentum along the
$\xi$-direction. 
The origin of oscillations in
these terms is quite clear physically. Qualitatively, these terms
are due to the interference between the photon emitted from the
$\xi =0$ brane towards negative  $\xi$ and  the
photon which was originally emitted with the positive momentum but
reflected from the second brane at $\xi=\xi_0$. 

The second term does not oscillate. This term 
is the one-loop correction to the cross-section of the
emission of the photon with  positive momentum along the
$\xi$-direction. 
It is
straightforward to see that this term exactly cancels the
cross-section of $e^+e^-$-annihilation into the $b^+b^-$-pair in the
physically interesting quantity --- the inclusive cross-section of
$e^+ e^-$-annihilation into the bulk. This cancelation is 
precisely what we have anticipated from
the  unitarity argument.

So we see that the physically interesting quantity --- inclusive
cross-section of
 $e^+e^-$-annihilation into the bulk receives  multiplicative
corrections due to  $b$-particles. The overall factor
differs from unity by a rapidly oscillating function of
energy; to the order $g^4$ one has 
\begin{eqnarray}
\sigma_{tot} (e^+ e^- \to {bulk})
= \sigma_0 (e^+ e^- \to \gamma_{{bulk}}) \times
\;\;\;\;\;\;\;\;\;\;\;\;\;\;\;\;\;\;\;\;\;\;\;\;
\;\;\;\;\;\;\;\;\;\;\;\;\;\;\;\;\;\;\;\;\;\;\;\;\; \nonumber \\
\l1-{g^2\cos(2E\xi_0)\over 16\pi}E+
{g^2\cos(2E\xi_0)\over16\pi^2}E\ln{E^2/\mu_{UV}^2}\r\;.
\end{eqnarray}
Similar result holds for other processes, for instance,
$e^+e^-$-annihilation into $e^+e^-$.

Let us discuss now what changes in the warped space, when the localized
photon mode is present. First of all, because of the localization, the
tree-level cross-section of $e^+e^-$-annihilation into the bulk photon 
is suppressed relative to the cross-section
of the annihilation into  $e^+e^-$ pair. Indeed,  the former
cross-section is equal to
\be
\label{RSeetogamma}
\sigma_0(e^+e^-\to\gamma_{{bulk}})={8\pi g^2k\over N_{\nu-1}(E/k)^2+
J_{\nu-1}(E/k)^2}{1\over E^2}
\ee
where
\[
  \nu = 1 + \frac{n}{2} \; ,
\]
while the latter has the standard four-dimensional
behavior at  $E\ll k$,
\be
\label{RSeetoee}
\sigma_0(e^+e^-\to e^+e^-)={\pi g^4\over 4}\left|{ H_\nu^{(1)}(E/k)
\over H_{\nu-1}^{(1)}(E/k) }\right|^2\approx {\pi g^4k^2\over E^2}(\nu-1)^2\;.
\ee
Consequently, assuming that  the effective
four-dimensional coupling constant $gk^{1/2}$ is not too small, 
we see that $e^+ e^-$-annihilation into a bulk photon
is suppressed by a factor
\be
\label{suppression}
\frac{\sigma_0(e^+e^-\to\gamma_{{bulk}})}{\sigma_0(e^+e^-\to e^+e^-)}
\equiv
\alpha(E/k)={1\over N_{\nu-1}(E/k)^2+
J_{\nu-1}(E/k)^2}\propto \l{E\over k}\r^{2\nu-2}\;.
\ee
Now, the 
photon propagator $G_A(0,\xi;p)$ from the 
$\xi=0$ brane to $\xi=\xi_0$ brane has
the 
following
form~\cite{katz} 
\be
\label{RSprop}
G_A(0,\xi_0;p)={i\over (2\pi)^4}{(1+k\xi_0)^\nu\over
2p}{H_{\nu}^{(1)}(p/k+p\xi_0) \over H_{\nu-1}^{(1)}(p/k)}\;.  
\ee 
This
propagator is again exponentially suppressed at relatively
large Euclidean
momenta 
\[
G_A \propto \mbox{e}^{-2|p|\xi_0} \; , \;\;\; |p| \gg \xi_0^{-1} \;
, \;\;\;
p^2 <0 \; ,
\]
so the contributions of $b$-particles into precision
QED observables are again exponentially suppressed at 
$\xi_0 \gg m_e^{-1}$. Furthermore,
the general argument based on the spectral
representation of the photon 
Green's function still applies to $e^+ e^-$-annihilation
into $b^+ b^-$. 
Let us elaborate on this argument a bit more.
In the Abelian theory, {\it the only} contribution
(besides $e^+ e^-$-loops) into the photon Euclidean
two-point
function, the left hand side of Eq.~(\ref{Challen}),
which is {\it not} suppressed by $\mbox{exp}({-2|p|\xi_0})$,
comes from the free photon propagator itself.
The corresponding contribution into the spectral density
$\rho(s)$ is due to the small tree-level cross section
$\sigma_0 (e^+ e^- \to \gamma_{bulk})$, see
Eq.~(\ref{suppression}). $b$-particle loops in the Euclidean two-point
function all involve at least two propagators (\ref{RSprop}), so they
all have exponential behavior (\ref{5a*}) in the Euclidean domain,
which corresponds to the oscillatory behavior (\ref{oscil})
of the spectral density. One immediately concludes that the inclusive
cross section of $e^+ e^-$-annihilation into bulk particles has the
form
\[
   \sigma_{tot} (e^+ e^- \to bulk) = \sigma_0 (e^+ e^- \to
   \gamma_{bulk})
+ (\mbox{oscillating~terms}) \; .
\]
Since this cross section is positive, the amplitude of oscillations
   should not exceed $\sigma_0 (e^+ e^- \to \gamma_{bulk})$.
We thus argue that the full inclusive cross section should be
suppressed, relative to $e^+ e^- \to e^+ e^-$, by the same factor
$(E/k)^{2\nu - 2}$  as the tree-level cross section of
$e^+ e^- \to \gamma_{bulk}$.

Let us see how diagrammatics for the cross section
of $e^+ e^-$-annihilation into bulk particles works
in the warped space. In this case the cross-sections
of $e^+e^-$-annihilation into $e^+e^-$- and $b^+b^-$-pairs are no
longer equal  to each other. Instead, their ratio is 
\be
\label{RSeetobb}
{\sigma(e^+e^-\to b^+b^-)\over\sigma(e^+e^-\to
e^+e^-)}=\left|{H^{(1)}_\nu(E/k+E\xi_0)\over
H^{(1)}_\nu(E/k)}\right|^2(1+k\xi_0)^{2\nu}
\ee 
At small $\xi_0$ this ratio approaches unity as it should be, while at
large values of $\xi_0$ it equals  
$(E\xi_0)^{2\nu}$ up to a numerical coefficient of order one. So,
naively,  the cross section of $e^+ e^-$-annihilation  
 into $b$-particles is even enhanced in warped space
relative to the annihilation  into the usual
electrons. However,  the observable quantity is, as before, 
the inclusive
cross-section of annihilation into the bulk.
To the order $g^4$, the latter is the
sum $\sigma_{tot}$
of the
cross-section (\ref{RSeetobb}) and one-loop improved cross-section 
$\sigma(e^+e^-\to\gamma_{{bulk}})$.
One finds that its ratio to the tree-level cross
section  (\ref{RSeetogamma}) is  
\begin{eqnarray}
\label{RSsum}
{\sigma_{{tot}}(e^+e^-\to {bulk})\over\sigma_0(e^+e^-\to\gamma)}=
1+{g^2k(1+k\xi_0)^{2\nu}\over32}\times\nonumber\\
\left[
f(E/k)\l N_\nu^2(E/k+E\xi_0)-J_\nu^2(E/k+E\xi_0)\r
+\nonumber\right.\\
\left.g(E/k)\l N_\nu(E/k+E\xi_0)J_\nu(E/k+E\xi_0)\r\right]
\end{eqnarray}
Here 
\begin{eqnarray}
f(E/k)={E^2\over k^2}\alpha(E/k)\l N_{\nu-1}(E/k)^2-J_{\nu-1}(E/k)^2
+\right.\nonumber\\
\left.
2\pi\ln{E^2/\mu_{UV}^2}
N_{\nu-1}(E/k)J_{\nu-1}(E/k)
\r\nonumber
\end{eqnarray}
and
\begin{eqnarray}
g(E/k)={E^2\over k^2}\alpha(E/k)\l 4N_{\nu-1}(E/k)J_{\nu-1}(E/k)
-\right.\nonumber\\
\left.2\pi\ln{E^2/\mu_{UV}^2}\l N_{\nu-1}(E/k)^2-J_{\nu-1}(E/k)^2\r
\r
\end{eqnarray}
are smooth functions of energy, which are both 
small at $E/k\ll 1$. We see that at
large $\xi_0$, the
inclusive cross-section of $e^+e^-$-annihilation into the bulk
particles
is
again modified by a multiplicative factor, which differs from
unity by a smooth function of energy times a rapidly oscillating
function,
\begin{eqnarray}
\label{generalcos}
{\sigma_{{tot, 1-loop}}(e^+e^-\to {bulk})\over\sigma_0(e^+e^-\to
\gamma_{{bulk}})} 
= \;\;\;\;\;\;\;\;\;\;\;\;\;\;\;\;\;\;\;
\;\;\;\;\;\;\;\;\;\;\;\;\;\;\;\;\;\;\;\;\;\;\;\;\;\;\;\;\;\;\;
\;\;\;\;\;\;\;\;\;\;\;\;
\nonumber \\
1-{g^2k(1+k\xi_0)^{2\nu}\over 16\pi E\xi_0}\l
f(E/k)\cos{2E\xi_0}+{1\over 2}g(E/k)\sin{2E\xi_0}
\r\;.
\end{eqnarray}
In spite of the smallness of functions $f(E/k)$ and $g(E/k)$,
the oscillating
terms in Eq.~(\ref{generalcos}) appear to be large due to
the fact that the effective coupling constant at the $\xi=\xi_0$
brane 
is equal to $gk^{1/2}(k\xi_0)^\nu$.
 Clearly, this is  an indication that one cannot
trust the perturbative calculations of the coefficients in front of
$\cos{2E\xi_0}$ and $\sin{2E\xi_0}$ in Eq.~(\ref{generalcos}). 
The unitarity argument together with diagrammatics suggest that
the complete cross section
has the general form
\begin{eqnarray}
\label{generalform}
\sigma_{{tot}}(e^+e^-\to
{bulk})=\;\;\;\;\;\;\;\;\;\;\;\;\;\;\;\;\;\;\;
\;\;\;\;\;\;\;\;\;\;\;\;\;\;\;\;\;\;\;\;\;\;\;\;\;\;\;\;\;\;\;
\;\;\;\;\;\;\;\;\;\;\;\;
\nonumber \\
\sigma_0(e^+e^-\to\gamma_{bulk}) \times \left(1+
f_r(E/k)\cos{2E\xi_0}+g_r(E/k)\sin{2E\xi_0} \right)
\end{eqnarray}
with coefficients $g_r$ and $f_r$ always smaller unity.
The suppression of
$e^+e^-$-annihilation into the bulk valid in the absence of bulk
fermions should hold after these fermions are taken into account.

Clearly, the general argument based on the spectral
representation applies not only to the two-point Green's function
(relevant to the annihilation processes) but to more general Green's
functions as well.
One may still 
worry that $b$-particles may dramatically enhance bremsstrahlung
of very soft photons,   $E \lesssim \xi_0^{-1} $, into the bulk. Such an
enhancement will certainly be avoided if $b$-particles are given a
mass $\mu \gg \xi_0^{-1}$.
%
This mass may be
quite small, so that electron decay is still possible, yet
$b$-particles decouple at energies below $\mu$, and do not
contribute into bulk photon bremsstrahlung at  $E \lesssim \xi_0^{-1}$.
A more elegant way to avoid the dangerous enhancement of bulk photon
bremsstrahlung is to consider
the genuine bulk field described by the action (\ref{ferm}) instead of
 $b$-particles living on the additional brane. 
Then, if one takes the Yukawa coupling $y$  large
enough,
\[
{yv\over k}\gg 1
\]
the wave function of a bulk fermion mode with four-dimensional mass $\mu$ is
unsuppressed
only at  $\xi \gg\mu^{-1}$, more precisely, at
\[
\xi \gtrsim {yv\over k\mu}\;.
\]  
The Euclidean photon
propagator with $|p|\gtrsim \mu$ is exponentially small in that region,
so  the above suppression mechanism  works for all bulk momenta.

It is worth noting that our general argument
{\it does not} work in the
{\it non-Abelian} case. Indeed, in that
case  self-interaction of vector fields   becomes
important precisely at the same values of $\xi$ where the 
Euclidean gauge field
propagator starts to decay exponentially. 
Hence, it is unclear whether the setup (\ref{*})
is suitable to localize non-Abelian gauge fields in a
phenomenologically acceptable way.
Perhaps, one can try to solve this
problem by tuning the profile of the warp factor.

To conclude, let us point out an analogy between 
multi-dimensional models discussed in this note
and purely four-dimensional models with charge non-conservation. 
The discussion of whether electric charge may not be exactly conserved
has  long history and the conclusion is that even tiny
non-conservation of electric charge leads to contradiction to
low-energy tests of QED unless millicharged particles are 
introduced~\cite{millicharged1,millicharged2,millicharged3,all} (see, however,
Ref.~\cite{however}). Millicharged particles appear ``naturally'' in
models with two massless Abelian gauge bosons $A^1_{\mu}$
and $A^2_{\mu}$ ~\cite{holdom}. In that case,
 mixing in kinetic term, $\epsilon
F_{\mu\nu}^{1}F_{\mu\nu}^{2}$, may be introduced.
To see that this term indeed gives rise to particles with small
(proportional to $\epsilon$) charges, it is convenient to work in the
basis where
photon $\gamma$ is a linear combination of $A^1$ and $A^2$ interacting with
ordinary quarks and leptons, and the second gauge boson $\gamma'$ 
(paraphoton) is 
another linear combination, such that the kinetic term for gauge fields
is diagonal. Then ``shadow'' matter charged under
$A^2$ gauge field interacts with physical photon
with strength proportional to $\epsilon$.

To see the link between the model with paraphoton and
multi-dimensional models, 
let us consider a different orthogonal
basis for gauge
fields in the former model, namely, introduce fields
\[
A^{\pm}=A^1\pm A^2\;.
\]
In this basis, the interaction of  the ``photon''
$A^+$ with shadow matter is of the same strength as its interaction
with  normal
matter (up to a correction of order  $\epsilon$). 
However, the interaction between shadow and normal matter is very weak
due to the screening effect of $A^-$. 
Now the analogy between the model with paraphoton and multi-dimensional
models is transparent --- the  photon zero mode plays a role 
of $A^+$,
while bulk photon modes collectively play a role of 
$A^-$-field\footnote{It is worth noting that the
 analogy is not complete, as the
$A^-$-field interacts at equal strength with normal and shadow
matter, while bulk photon modes interact  weakly with localized
matter and  strongly with bulk (``shadow'') matter. A more complete
analogy may be obtained if one considers $A^+$-photon interacting with
charge $e$ with all types of matter, and $A^-$-photon interacting with
 charge $e/N$ with normal matter and with charge $Ne$ with shadow
matter ($N\gg1$).}. Analogs of millicharged particles are the
bulk modes of  electron.

This analogy stresses the point that
 the real physical photon created by brane matter
is not a bare Kaluza--Klein zero mode but a certain energy-dependent
admixture of the zero
mode and bulk modes. 
There is one more aspect of this analogy: a holographic interpretation
 of the brane image of a charged particle moving in the bulk would be
 a charged cloud of (quasi-)conformal matter. This cloud has a clear
 analogy to a cloud of millicharged particles, into which an electron
would decay in paraphoton models.

We are indebted to M.~Libanov, J.~Polchinski, R.~Rattazzi,
P.~Tinyakov, N.~Seiberg and M.~Shaposhnikov for useful
discussions. We thank the organizers of Francqui
Meeting on Strings and Gravity and
XXXVII Rencontres de Moriond for warm and stimulating
atmosphere. This work has been supported in part by
Russian Foundation for Basic Research, grant 02-02-17398,
and Swiss Science Foundation grant 7SUPJ062239.
The work of S.D. has been supported in part 
by the INTAS grant YS 2001-2/128.

\end{document}